\documentclass[12pt]{article}
 \usepackage{epsfig}
 \usepackage{amsmath}
 \def\be{\begin{equation}}
 \def\ee{\end{equation}}
 \def\bea{\begin{eqnarray}}
 \def\eea{\end{eqnarray}}
 \usepackage{subcaption}
 \usepackage{graphicx}

 \usepackage[symbol]{footmisc}

 \catcode`\@=11
 \def\lsim{\mathrel{\mathpalette\@versim<}}
 \def\gsim{\mathrel{\mathpalette\@versim>}}
 \def\@versim#1#2{\vcenter{\offinterlineskip
 \ialign{$\m@th#1\hfil##\hfil$\crcr#2\crcr\sim\crcr } }}
 \catcode`\@=12

 \catcode`@=12
 \def\R{{\cal R}}
 \def\H{{\cal H}}
 \def\C{{$\cal~ C$}}
 \def\c{{\cal C}}
 
 \def\CP{{$\cal CP~$}}
 \def\B{{\cal B}}

 \def\KS{{|K_S\rangle}}
 \def\KL{{|K_L\rangle}}
 \def\KO{{|K_1\rangle}}
 \def\KT{{|K_2\rangle}}
\def\Ka{{|K_a\rangle}}
\def\Kb{{|K_b\rangle}}
 
\begin{document}
 \thispagestyle{empty}
 \mbox{}
 \vskip .2in 
 \begin{center}
 {\LARGE \bf Geometric Phases in Kaon Decays and Baryogenesis}
 \vskip 0.8in
 {\bf Swarup Sangiri$^1$\footnote[1]{swarup.phys@gmail.com} and Utpal Sarkar$^2$ \\}
 \vspace{0.2in}
{\sl $^1$ Physics Department,
 Indian Institute of Technology, Kharagpur 721302,  India \\}
\vspace{0.1in}
{\sl $^2$ Department of Physical Sciences, \\ Indian Institute of 
	Science Education and Research Kolkata, Mohanpur 741246, India \\}
 \end{center}
 \vspace{0.8in}

\begin{abstract}\
	We studied the formalism for construction of Bargmann invariants
(BIs) as quantum mechanical geometric phases and identified the 
\CP invarince with the rephasing invariant  phases in 
neutral kaon system, kaon decays, baryogenesis and leptogenesis.
We develop this formalism to express the \CP violation in terms
of the Bargmann invariants, which allow us to interpret them as
geometric phases. We then comment on the 
application of such generalized treatment of \CP phases. 
\end{abstract}

\newpage
\baselineskip 24pt


\section{Introduction}
	
	The product of the charge conjugation and the parity, denoted as \CP, plays a crucial role in
	particle physics. \CP violation has been observed in kaon 
	decays \cite{christensen} and \CP violation is needed to explain the matter antimatter
	asymmetry of the universe \cite{sakh}. Thus understanding the \CP violation in any theory is very important, so while
	proposing any new extensions of the standard model, it is imperative to
	study \CP violation in such models.
	We thus envisage a generalized treatment of \CP violation that may be applied
	to the different formalisms and allow them to compare with each other. For this purpose, 
	we propose to adopt the construction of the
	quantum mechanical  Bargmann invariants and apply them to the
	different scenarios of \CP violation. In this article we shall study the
	\CP violation in $K^0 -\overline{K^0}$ oscillation, K-meson decays,
	baryogenesis,  
	leptogenesis with right-handed neutrinos and leptogenesis with
	triplet Higgs scalar and construct the Bargmann invariants
	in each case, demonstrating the
	geometric nature of the measure of  \CP violation.  
	
	The kaon decays have been studied extensively in the literature and this is the
	first place where \CP violation was experimentally observed \cite{christensen}. There are two sources
	of \CP violation in the kaon decay, usually denoted by $\epsilon$ and 
	$\epsilon^\prime \slash \epsilon$; where $\epsilon$ represents the \CP violation
	in $\Delta S = 2$ transition of $K^\circ - \overline{K^\circ} $ 
	oscillation and $\epsilon^\prime \slash \epsilon$ represents the \CP violation in
	direct neutral kaon decays which is a $\Delta S =1$ process \cite{bigi}. $S$ is the strangeness
	of $\overline{K^\circ}$. Usually the $\epsilon$ appears in the mass matrix of 
	the neutral kaons and relates the \CP eigenstates to the physical mass eigenstates
	of the kaons. At the quark level, the $K^\circ - \overline{K^\circ} $
	oscillation comes from a box diagram, but we shall restrict ourselves
	to the meson states, which are experimentally observed, where it will emerge
	from a self-energy diagram involving kaons and pions. On the other hand, 
	$\epsilon^\prime \slash \epsilon$ appears in the penguin diagram at the
	quark level, which comes from the usual vertex correction at the meson
	level.

	In this article we shall study the \CP violation in the kaon system without
	going into the quark model. Since quarks are not observable and only
	kaons and their decay products, the pions, are observables, this approach allow us to interpret the results and relate the \CP violating parameters to the quantum mechanical geometric
	phases, which in turn, relates them to the Bargmann invariants consistently. 
	We shall then demonstrate 
	how the \CP violation appearing in the process of baryogenesis and leptogenesis may be
	related to the Bargmann invariants in the kaon decays. In both cases the
	parameter $\epsilon$ comes from a self energy diagram, while the  $\epsilon^\prime \slash \epsilon$ 
	appears in the vertex  diagrams. 

The geometric phase was originally discovered in the context of cyclic adiabatic quantum  mechanical evolution, govorned by the time-dependent Schrodinger equation with a hermitian Hamiltonian operator \cite{berry}. Subsequent generalization removed many of these restrictions and geometric phase has been defined in nonadiabatic \cite{Aha}, noncyclic and nonunitary evolution \cite{sam}. In the process of our understanding of the geometric phase has brought together many aspects of the basic structure of quantum mechanics \cite{MukSim}. Geometric phases are simple and elegent at the conceptual level, but it captures the essence
of foundations of quantum mechanics. The key concept behind this formalism is 
parallel transport of vectors and the inner products relating the vector spaces
with their dual vector space. 
\

Bargmann invariants \cite{bargmann} are constructed using  inner products of vectors, which we shall define in the following sections. With some of the  generalizations stated above, a connection between geometric phases and Bargmann invariants were established \cite{BI-GP},\cite{BI-GP1}. For a physical system, the state vectors represents Dirac fermions, and the Bargmann invariants may be identified with the rephasing invariant measures of \CP violation. This result has been generalised for the case when the state vectors represent both Dirac and Majorana fermions by defining the ray and the Hilbert spaces properly \cite{BI-majorana}. This introduced additional Bargmann invariants representing \CP violation arrising from the Majorana phases. Although the existence of Majorana fermions has not yet been established, many interesting aspects of the Majorana fermions have been pointed out both in particle physics and condensed matter physics \cite{majorana}. There have been extensive studies about Majorana fermions in topological superconductors where they can be constructed using quasiparticle excitations \cite{cwj}. There have also been rigorous studies of geometric phase in various fields including condensed matter \cite{resta}. Thus, constructing BIs for Majorana fermions in condensed matter and hence establish a connection with geometric phases remains an exciting field, on which we intend to work in the future.
\
\section{\CP violation, Bargmann Invariants and Geometric Phases}

\CP violation has been studied extensively, but still there are many issues that
need to be understood both theoretically and experimentally. A connection between
the geometric phases and the \CP violation have been studied in the  case of Dirac fermions \cite{BI-GP} and have been extended to include Majorana fermions.

In this section we shall summarize the formalism of Bargmann invariants (BIs) and
their connection to the quantum mechanical geometric phases with the \CP
violation in the system. We shall follow the formalism and notation of \cite{BI-GP},\cite{BI-GP1}.

Let $\H$ be a complex Hilbert space consisting of vectors $\psi,\phi,...$ and $\R$  be the associated ray space. $\H$ contains both Dirac and Mjorana particles. The subset of unit vectors in $\H$ is defined by 
\begin{equation}
\B = \{\psi \in \H ~|~ (\psi,\psi)=1\}\subset \H,
\end{equation}
where (\ ,\ ) denotes the inner product between two states. Ray space $\R$ is the quotient of $\B$ with respect to the U(1) action, i.e, with respect to the equivalence relation $\psi \sim e^{i\alpha} \psi$ for all real phases $ \alpha$. Elements of $\R$ are represented by projection operators onto unit vectors:
\begin{equation}
\R=\{\rho(\psi)=\psi \psi^{\dagger} ~|~\psi\in\B \}
\end{equation}
The projection from $\B$ to $\R$ is denoted by
\begin{eqnarray}
\pi:\B\rightarrow\R:\psi\in \B\rightarrow\rho(\psi)=\psi\psi^{\dagger} \in\R
\end{eqnarray}
We describe continuous parametrized curves $\c\subset\B$, and their projections $C\subset \R$ as 
\begin{equation}
\c=\{\psi(s)\in \B ~|~s_1\le s \le s_2  \}\subset \B,
\end{equation}
\begin{equation}
C=\pi(\c)=\{\rho(s)=\psi(s)\psi(s)^{\dagger}\in \R~|~ s_1\le s\le s_2\}  \subset \R.
\end{equation}
Now, any curve $\c$ in $\H$, projecting on to a curve $C$ in $\R$, is called a lift of the later. The most general lift is described as:
\begin{eqnarray}
\c^{\prime}=\{\psi^{\prime}(s) =e^{i\alpha(s)}\psi(s)~|~\psi(s)\in \c, s_1\le s\le s_2\}\subset \B
\end{eqnarray} 

In particular we have a horizontal lift $\c^{(h)}$ such that:

\begin{equation}
\left( \psi^{(h)}(s), ~\frac {d}{ds} {\psi}^{(h)}(s) 
\right)=0\,.
\end{equation}
The geometric phase can be defined for any curve $C\subset \R$, and can be calculated using any lift $\c$:
\begin{eqnarray}
\phi_{geo}[C]=\phi_{tot}[\c]-\phi_{dyn}[\c]\nonumber,  \\
\phi_{tot}[\c]=arg(\psi(s_1),\psi(s_2))\nonumber, \\
\phi_{dyn}[\c]=Im \int_{s_1}^{s_2} {ds \left(\psi(s),\frac{d}{ds}{\psi(s)}\right )}
\end{eqnarray}

If $\c$ is horizontal, $\phi_g[C]=\phi_{tot}[C]$

$\phi_{geo}[C]$ is both reparameterization and gauge invariant.
For any parameterized curve in ray space, we associate a length functional:
\begin{equation}
L[C]=\int_{s_1}^{s_2}\{(\dot{\psi}(s),\dot{\psi}(s))-(\psi(s),\dot{\psi}(s))(\dot{\psi}(s),\psi(s))\},
\end{equation}
Here $\dot{\psi}(s)=\frac{d}{ds}{\psi(s)}$.
Note that, $L$ does not depend on any lift $\c$, it only depends on $C$. Also $L$ is reparameterization invariant.  In ray space $\R$, free geodesics are those curves for which $L[C]$ is minimum for given end points. Free geodesics in $\H$ are any lift $\c$ of a free geodesic in $\R$. Now it can be shown that 
\begin{equation}
\phi_g[free\  geodesic\  in\  \R]=0
\end{equation}

This property is used to connect geometric phases to Bargmann invariants.

\
Bargmann invariants are quantum mechanical expressions defined on $\R$ involving inner products of states living in $\c$ in a cyclic manner. The order of the Bargmann invariant(BI) indicates the number of  inner products in it's expression. The first order BI of a vector $\psi$ living in $\B$ is nothing but the inner product of $\psi$ with itself : $(\psi,\psi)=1$. The second order BI for two states $\psi_1,\psi_2 \in \B$ is expressed as :
\begin{eqnarray}
\Delta_2(\psi_1,\psi_2)=(\psi_1,\psi_2)(\psi_2,\psi_1)=Tr(\rho_1\rho_2),
\end{eqnarray}
which is real, non-negative.
To get something more interesting, we write the third order Bargmann invariant :
\begin{equation}
\Delta_3(\psi_1,\psi_2,\psi_3)=(\psi_1,\psi_2)(\psi_2,\psi_3)(\psi_3,\psi_1)=Tr(\rho_1 \rho_2\rho_3),
\end{equation}
where $\psi_1$, $\psi_2$, $\psi_3$ are three vectors in $\B$, with no two $\psi$ mutually orthogonal. $\rho_1$, $\rho_2$, $\rho_3$ are the corresponding images in ray space $\R$. This third order BI is in general complex.
BIs are ray space quantity, because it remains invariant under gauge transformation.
The m-sided BI is written as
\begin{equation}
\Delta_m(\psi_1,\psi_2,...,\psi_m)=(\psi_1,\psi_2)(\psi_2,\psi_3)...(\psi_m,\psi_1)=Tr(\rho_1\rho_2...\rho_m)
\end{equation}

The connection between BI and free geodesics is represented in the following way: Consider an m-sided polygon in $\R$ whose sides are free geodesics connecting $\rho_1$ to $\rho_2$, $\rho_2$ to $\rho_3$,...,$\rho_m$ to $\rho_1$. Then using the result (10) it was established that: 
\begin{multline}
\phi_g[n-vertex \ polygon\ in\ \R\ connecting \ \rho_ 1 to \ \rho_2,\ \\ \rho_2\ to\ \rho_3,..., \rho_m\ to\ \rho_1\ by\ free\ geodesics]= -arg\Delta_m(\psi_1,\psi_2,...,\psi_m)
\end{multline}

As long as any successive pair of vectors are not mutually orthogonal, we obtain a BI with well defined phase.

\section{Bargmann invariants in the neutral kaon system }
The connection between Bargmann invariants and the rephasing invariant quantities with the geometric phase for the Dirac fermions has been established. This connection has been further extended to include Majorana fermions by relationg Bargmann invariants with the Majorana phases. Here we shall develop the connection between the \CP violating parameters for neutral kaons with the Bargmann invariants. \CP violation in the $K^0-\bar{K^0}$ oscillation is a $\Delta S=2$ process, changing the strangeness quantum number S by two units. This process is also called indirect \CP violation. The \CP violating parameter $ \epsilon $ comes through the mass matrix of the $K^0-\bar{K^0}$ system. In the quark sector the dominant contribution of the $K^0-\bar{K^0}$ oscillation comes from a box diagram, and \CP violation enters through a quark mixing matrix. But here, rather than going to the quark model, we use the effective fields ($K$ and $\pi$ mesons)  to establish the connection between \CP violation in neutral kaon decay and geometric phases.

\
The physical $K$ meson states, or the mass eigenstates of neutral kaons are represented by orthonormal basis $|K^0\rangle,\bar{|K^0\rangle}$. The short and long lived kaons are represented by nonorthogonal basis $|K_S\rangle,|K_L\rangle$ with the following relation 
\begin{eqnarray}
|K_S\rangle={1\over{\sqrt{2(1+|\epsilon|^2)}}}~(~(1+\epsilon)|K^0\rangle+(1-\epsilon)|\bar{K^0})~)~\nonumber \\
|K_L\rangle={1\over{\sqrt{2(1+|\epsilon|^2)}}}~(~(1+\epsilon)|K^0\rangle-(1-\epsilon)|\bar{K^0})~)~
\end{eqnarray}
where $\epsilon$ is the \CP-violating complex phase. Now $K^0$ and $\bar{K^0}$ can be expressed in terms of \CP eigenstates as 
\begin{eqnarray}
|K^0\rangle={1\over \sqrt{2}}[\KO +\KT] \nonumber
\\
|\bar{K^0}\rangle={1\over \sqrt{2}}[\KO -\KT] 
\end{eqnarray}
After substituting these values we find :
\begin{eqnarray}
\KS={1\over \sqrt{1+|\epsilon|^2}}[\KO+\epsilon \KT]\nonumber \\
\KL={1\over \sqrt{1+|\epsilon|^2}}[\KT+\epsilon \KO]
\end{eqnarray}
For constructing the BIs, we start with the state vectors $\KS$, $\KL$, $\KO$ and $\KT$ in the Hilbert space $\mathcal{H}$. We use the orthogonality of the $K_{1,2}$ basis
\begin{eqnarray}
(K_i,K_j)=\delta_{ij},
\end{eqnarray}
where $i=1,2$ and $j=1,2$, and using Eq. 17, we express the inner products of the non-orthogonal states as
\begin{eqnarray}
\begin{aligned}[b]
(K_S,K_L)={2\over ({1+|\epsilon|^2})}Re(\epsilon) \\
(K_S,K_1)=(K_L,K_2)={1 \over \sqrt{ ({1+|\epsilon|^2})}} \\
(K_S,K_2)=(K_L,K_1)={\epsilon^* \over \sqrt{({1+|\epsilon|^2})}}.
\end{aligned}
\end{eqnarray}
We can now construct the following two BIs using the inner products of $K_S, K_L$ and \CP eigenstates $K_1,K_2$:
\begin{eqnarray}
\Delta_3(K_L,K_S,K_1)=(K_L,K_S)(K_S,K_1)(K_1,K_L)={2\epsilon Re\{ \epsilon\}\over (1+|\epsilon|^2)^2}
\\
\Delta_3(K_L,K_S,K_2)=(K_L,K_S)(K_S,K_2)(K_2,K_L)={2\epsilon^* Re\{ \epsilon\}\over (1+|\epsilon|^2)^2}
\end{eqnarray}

Adding Eq. (20) and Eq. (21):
\begin{eqnarray}
\Delta_3(K_L,K_S,K_1)+\Delta_3(K_L,K_S,K_2)={4( Re\{ \epsilon\})^2\over (1+|\epsilon|^2)^2}
\end{eqnarray}

Now \CP violating phase $\delta$, which is physically observable, 
is defined as:
\begin{eqnarray}
\delta ={\Gamma(K_L\rightarrow \pi^- l^+ \nu)-\Gamma(K_L\rightarrow\pi^+ l^-\bar{\nu})\over \Gamma(K_L\rightarrow \pi^- l^ + \nu)+\Gamma(K_L\rightarrow\pi^+ l^-\bar{\nu})} = {2Re\{\epsilon\}\}\over 1+|\epsilon|^2}
\end{eqnarray}

So we have the following result 
\begin{eqnarray}
{\Delta_3(K_L,K_S,K_1)+\Delta_3(K_L,K_S,K_2)=\delta^2} 
\end{eqnarray}

This is one of the key results of our work. The Bargmann invariants are defined in the ray space $\R$, and the points $\rho(K_1),\ \rho_(K_2),\ \rho(K_L)$ etc on $\R$ are connected through a closed loop forming a geodesic. Now geometric phase can be obtained from BIs : $\phi_g=-arg(\Delta)$. Hence Eq. 24 represents the connection between a combination of geometric phases and \CP violation.

Here we have only emphasized on third order BIs. We can construct fourth order BIs with the states  $\KS$, $\KL$, $\KO$ and $\KT$ in the Hilbert space $\mathcal{H}$. Using the orthogonality of $K_{1,2}$ states and Eq. 17 we can obtain nonzero BIs $\Delta_4(K_S,K_1,K_L,K_2)$ and $\Delta_4(K_S,K_2,K_L,K_1)$. But we do not obtain any new results constructing them, because these fourth order BIs have a linear relationship with the third order BIs we already obtained:
\begin{eqnarray}
\begin{aligned}[b]
\Delta_4(K_S,K_1,K_L,K_2)-\Delta_4(K_S,K_2,K_L,K_1)
&=\Delta_3(K_L,K_S,K_1)-\Delta_3(K_L,K_S,K_2)
\end{aligned}
\end{eqnarray}
Hence it is enough to work only with the third order BIs for the case of neutral kaon oscillations. Fourth order BIs will naturally appear for the case of \CP violation in neutral kaon decay, as we shall discuss in the next section.

\section{Connection between Bargmann invariants and neutral kaon decays}
Unlike neutral kaon oscillation, kaons decaying into pions is a $|\Delta S|=1$ process, which changes the strangeness by one unit, known as direct \CP violation. The contribution of $\epsilon^\prime$ comes from the $|\Delta S|=1$ process. In the quark model, the dominant contribution to $\epsilon^\prime$ comes from a gluon mediated penguin diagram. The vertex correction in neutral kaon decay contributes to both $|\epsilon^\prime/\epsilon|$ and $\epsilon$. Both direct and indirect \CP violation enters in the CKM mixing matrix through a \CP phase \cite{Hocker}.

In the quark level, considering Yukawa coupling between fermions and scalars, a rephasing invariant has been defined, and it's connection with \CP violation has been established \cite{green}. This invariant is called the Jarlskog invariant \cite{jarlskog}. In this section we shall establish a connection between the Bargmann invariants with the neutral kaon decays. The BIs we shall obtain will be shown to be equal to rephasing invariants for the case of neutral kaons decaying to pions. As stated in the previous section, our formalism shall be restricted to scalar sector and we shall see that we can obtain rephasing invariants and their connection with the BIs without going to the quark level.  

Kaons can decay to two pion final state with $I=0$ or $2$, because Bose statistics constrain the $2\pi$ system to carry isospin 0 or 2. In this section we shall denote $\KS$ and $\KL$ as $\Ka$ and $\Kb$ respectively, and $|(\pi\pi)_0\rangle $ and $(\pi\pi)_2\rangle $ as $|(\pi\pi)_i\rangle $ and $|(\pi\pi)_j\rangle $ respectively. The states $\Ka$ and $\Kb$ decays to two pion final states as:
\begin{eqnarray}
\Ka=f_{ai}|(\pi\pi)_i\rangle+f_{aj}|(\pi\pi)_j\rangle  \nonumber \\
\Kb=f_{bi}|(\pi\pi)_i\rangle+f_{bj}|(\pi\pi)_j\rangle,
\end{eqnarray}
where $f$'s are the coupling strengths. 

\begin{figure}[h!]
	\centering
	\begin{subfigure}[b]{0.4\textwidth}
		\centering
		\includegraphics[width=\textwidth]{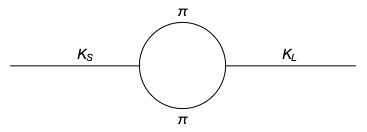}
		\caption{I=2}
		\label{}
	\end{subfigure}
	\hfill
	\begin{subfigure}[b]{0.4\textwidth}
		\centering
	\includegraphics[width=\textwidth]{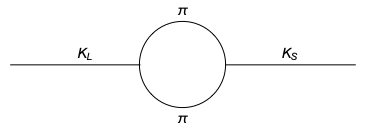}
		\caption{I=0}
		\label{}
	\end{subfigure}
	
	\caption{Mixing of $K_S$ and $K_L$ through pions}
	\label{self_energy_kaons}
\end{figure}

We start with the states $\Ka$, $\Kb$, $(\pi\pi)_i$ and $(\pi\pi)_j$ in the Hilbert space $\mathcal{H}$. We note that $\Ka$ and $\Kb$ are normalized, but not orthogonal:
\begin{eqnarray}
(K_a,K_b)\neq 0 \nonumber \\
(K_a,K_a)=(K_b,K_b)=1.
\end{eqnarray}
This allows us to express the inner products of the vectors as
\begin{eqnarray}
(K_\alpha,(\pi\pi)_\beta)=f^*_{\alpha\beta},
\end{eqnarray} 
where $\alpha=a,b$ and $\beta=i,j$. 
Now to find BIs and to establish a connection between them and rephasing invariant for the case of neutral kaon decay, we first consider the self energy diagrams of fig (\ref{self_energy_kaons}).
\

These two diagrams individually do not imply any \CP violation, because the amplitude comes out to be real. But if we consider the interference term from the two processes $K_S\rightarrow(\pi\pi)_2\rightarrow K_L$ followed by $K_L\rightarrow(\pi\pi)_0\rightarrow K_S$ we get the following term :
\begin{eqnarray}
f_{aj}^*f_{bj}f_{bi}^*f_{ai}I_{ab}
\end{eqnarray}

where $I_{ab}$ is the phase factor and $f_{ai}$ signifies the process $K_S\rightarrow (\pi\pi)_0$ and so on. The term $f_{aj}^*f_{bj}f_{bi}^*f_{ai}$ is rephasing invariant, in other words this term in invariant under all of the following phase changes

\centerline {$\Ka\rightarrow e^{i\delta_a}\Ka $}

\centerline{$\Kb\rightarrow e^{i\delta_b}\Kb$}

\centerline {$|(\pi\pi)_i\rangle \rightarrow e^{i\delta_i}|(\pi\pi)_i\rangle $}

\centerline {$|(\pi\pi)_j\rangle \rightarrow e^{i\delta_j}|(\pi\pi)_j\rangle$}

Now we can express this rephasing invariant quantity in terms of BIs. To construct the BI, we again consider the Hilbert space $\H$ which contains the states $\Ka, \Kb$, $|(\pi\pi)_i\rangle, |(\pi\pi)_j\rangle$. Now we consider any lift into ray space $\B$ and define the following BI in the Ray space
\begin{eqnarray}
\begin{aligned}[b]
\Delta_4(K_a, (\pi\pi)_j, K_b, (\pi\pi)_i)&=(K_a,(\pi\pi)_j)((\pi\pi)_j,K_b)(K_b,(\pi\pi)_i)((\pi\pi)_i,K_a) \\
&=Tr\ [\rho(K_a)\rho((\pi\pi)_j)\rho(K_b),\rho((\pi\pi)_i)] \\
&=Tr[(K_aK_a^\dagger)((\pi\pi)_j)(\pi\pi)_j^\dagger)(K_bK_b^\dagger)((\pi\pi)_i)(\pi\pi)_i)^\dagger] \\
&=f_{aj}^*f_{bj}f_{bi}^*f_{ai}.
\end{aligned}
\end{eqnarray}

This is another key result of our work, where we developed the connection between the rephasing invariant quantity which comes from the interaction term of self energy diagrams of neutral Kaon decaying into $I=0$ and $I=2$ with a purely geometrical quantity, the BI.

\

Now we shall consider one loop corrections to the tree level of neutral kaon decay.
\\
\begin{figure}[h!]
	\centering
	\begin{subfigure}[b]{0.4\textwidth}
		\centering
 	\includegraphics[width=\textwidth]{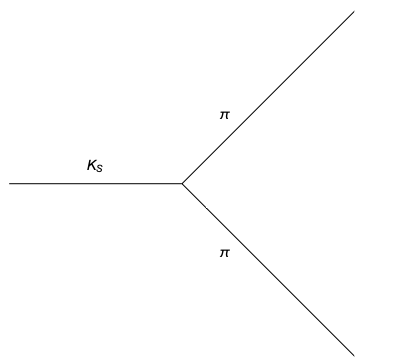}
		\caption{}
		\label{}
	\end{subfigure}
	\hfill
	\begin{subfigure}[b]{0.4\textwidth}
		\centering
		\includegraphics[width=\textwidth]{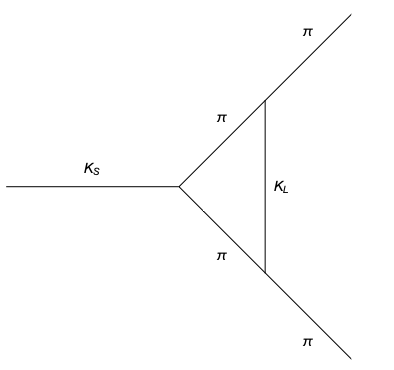}
		\caption{}
		\label{}
	\end{subfigure}
	
	\caption{Interaction between (a) tree level diagram and (b) one loop 
		vertex correction for the decay of neutral Kaons into two pions}
	\label{}
\end{figure}
The lowest order diagram gives a matrix element which is real. So we 
consider the one loop correction which gives a non-real interaction term.
After calculating the interaction term of these figures we can obtain the 
only rephasing invariant quantity 
\\
\centerline{${K_{ab}} f_{ai}^*f_{bi}f_{bj}^*f_{aj}$}
\
where, $K_{ab}$ is another phase constant.
We can establish the following relation between the BI and the rephasing invariant quantity 

\begin{eqnarray}
\begin{aligned}[b]
\Delta_4(K_a, (\pi\pi)_i, K_b, (\pi\pi)_j)&=(K_a,(\pi\pi)_i)((\pi\pi)_i,K_b)(K_b,(\pi\pi)_j)((\pi\pi)_j,K_a)\\
&=Tr\ [\rho(K_a)\rho((\pi\pi)_i)\rho(K_b),\rho((\pi\pi)_j)] \\
&=Tr[(K_aK_a^\dagger)((\pi\pi)_i)(\pi\pi)_i^\dagger)(K_bK_b^\dagger)((\pi\pi)_j)(\pi\pi)_j)^\dagger] \\
&=f_{ai}^*f_{bi}f_{bj}^*f_{aj}
\end{aligned}
\end{eqnarray}

The Bargmann invariants are defined on $\R$ connecting the points $\rho(K_a),\rho((\pi\pi)_i),\rho(K_b)$ and $\rho((\pi\pi)_j)$, forming a closed loop on $\R$. Also the connecting points mentioned above are non-orthogonal and pair wise linearly independent making the closed loop connecting them a geodesic. Thus the BIs, and in turn the rephasing invariant measures, gives the geometric phase 
\begin{eqnarray}
\phi_g=-arg(\Delta_4)
\end{eqnarray}
Hence, $\Delta_4$ is the rephasing invariant measure of \CP violation in  neutral kaon system. We obtained both third and fourth order BIs for the case of neutral kaons. In section 3 third order BI appeard while connecting \CP violation for kaon system and geometric phases, whereas in this section fourth order BI appeared while connecting rephasing invariants with the geometric phases. 
\section{Bargmann invariants in the theory of Baryogenesis}
The excess of baryons over anti-baryons, from the initial condition $B=0$, can be generated through baryon number violating decays of very heavy particles \cite{kolb}. Sakharov showed that in addition to the Baryon number violation, this process also requires both \C\  and \CP violation and the process must happen during a non thermal equilibrium state of the universe \cite{sakhrov}. Thus any theory of Baryogenesis must incorporate \  \CP violation. In this section we shall show the connection between \CP violating parameter in baryogenesis from heavy particle decay with the Bargmann invariants, and present the similarity of this connection with what we have established in the previous sections for the decay of neutral kaons.
\
\begin{figure}[h!]
	\centering
	\begin{subfigure}[b]{0.35\textwidth}
		\centering
		\includegraphics[width=\textwidth]{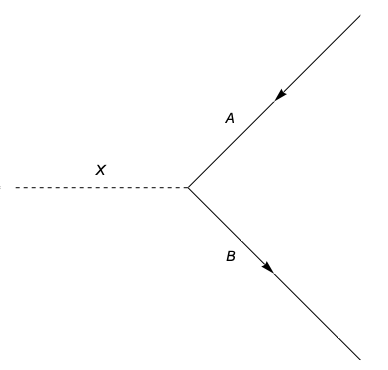}
		\caption{}
		\label{}
	\end{subfigure}
	\hfill
	\begin{subfigure}[b]{0.35\textwidth}
		\centering
		\includegraphics[width=\textwidth]{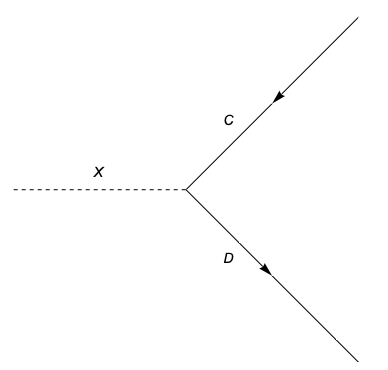}
		\caption{}
		\label{}
	\end{subfigure}
	\hfill
	\begin{subfigure}[b]{0.35\textwidth}
		\centering
		\includegraphics[width=\textwidth]{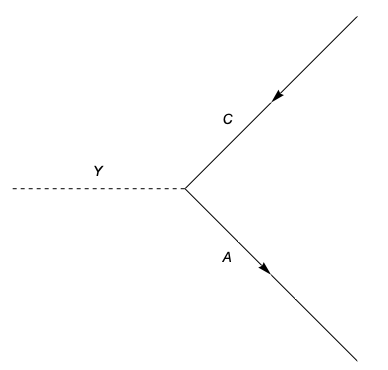}
		\caption{}
		\label{}
	\end{subfigure}
	\hfill
	\begin{subfigure}[b]{0.35\textwidth}
		\centering
		\includegraphics[width=\textwidth]{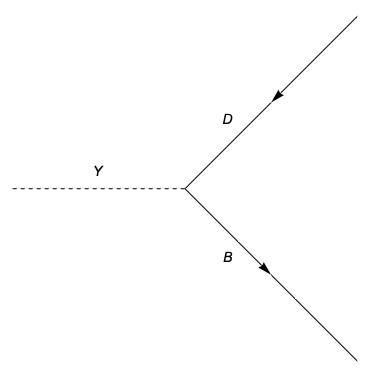}
		\caption{}
		\label{}
	\end{subfigure}
	
	\caption{Lowest order decays of X and Y into fermions}
	\label{XY_lowest_order}
\end{figure}

To illustrate this parallel, we shall consider the decay of two massive bosons X and Y. The interaction of X and Y with the standard model fermions A,B,C and D is given by the Yukawa couplings: 
\begin{eqnarray}
\mathcal{L}=f_{1ab^*}B^{\dagger}AX+f_{1cd^*}D^{\dagger}CX+f_{2a^*c}A^{\dagger}CY+f_{2b^*d}B^{\dagger}DY+h.c.
\end{eqnarray}
where $f$s are coupling strengths. 

\begin{figure}[h!]
	\centering
	\begin{subfigure}[b]{0.35\textwidth}
		\centering
	\includegraphics[width=\textwidth]{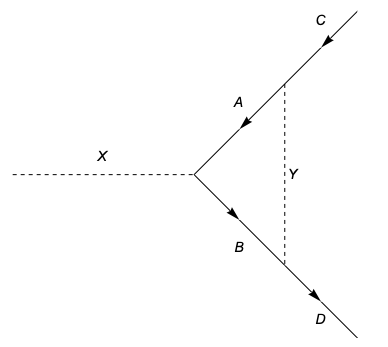}
		\caption{}
		\label{}
	\end{subfigure}
	\hfill
	\begin{subfigure}[b]{0.35\textwidth}
		\centering
		\includegraphics[width=\textwidth]{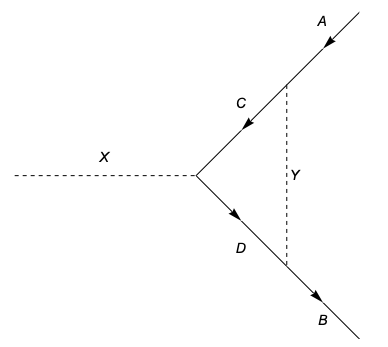}
		\caption{}
		\label{}
	\end{subfigure}
	\hfill
	\begin{subfigure}[b]{0.35\textwidth}
		\centering
		\includegraphics[width=\textwidth]{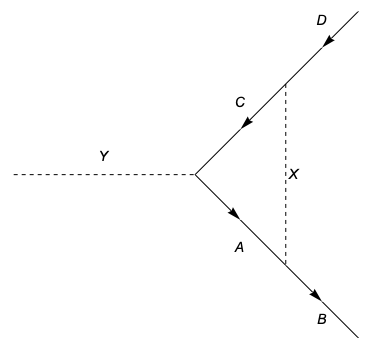}
		\caption{}
		\label{}
	\end{subfigure}
	\hfill
	\begin{subfigure}[b]{0.35\textwidth}
		\centering
		\includegraphics[width=\textwidth]{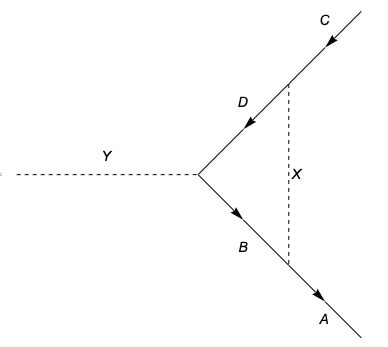}
		\caption{}
		\label{}
	\end{subfigure}
	
	\caption{One loop corrections for the decays of X and Y}
	\label{one_loop}
\end{figure}

Fig (\ref{XY_lowest_order}) shows the lowest order decay processes 
$X\rightarrow \bar{A}B,\bar{C}D$ and $Y\rightarrow \bar{C}A,\bar{D}B$. 
As the lowest order processes can not contribute to \CP violation, we 
consider the interference of this with the one loop corrections shown 
in fig (\ref{one_loop}). 
The rephasing invariant parameter of this interference is given by 
\begin{eqnarray}
\delta_B=(f_{1ab^*})^*f_{1cd^*}(f_{2a^*c})^*f_{2b^*d}\nonumber\\
=f^*_{1a^*b}f_{1cd^*}f^*_{2ac^*}f_{2b^*d}.
\end{eqnarray}
$\delta_B$ remains invariant under the phase changes of $|X\rangle$, $|Y\rangle$,  $|A\rangle$, $|B\rangle$, $|C\rangle$ and $|D\rangle$. 
 This quantity is equivalent to the Jarlskog invariant for \CP violation which appears in the quark mass matrix. 
 Now $Im(\delta_B)$ appears in the expression of baryon asymmetry in the XY model as:
 \begin{eqnarray}
 (\Delta B)\propto Im(\delta_B),
 \end{eqnarray}
 where $\Delta B=(\Delta B)_X+(\Delta B)_Y$ is the net baryon number produced in the decays of X and Y. $\Delta B$ vanishes if \CP is conserved. 
   
  We shall now construct Bargmann invariants for the XY model of baryogenesis and establish a relation between $\delta_B$ and the Bargmann invariants. Here we shall follow a close parallel with the neutral Kaon decay to establish such relations. We consider the dacay of the heavy boson X. Let us define $|R_1\rangle =|B^\dagger A\rangle $ and $|R_2\rangle =|D^\dagger C\rangle $, for convenience. $|R_1\rangle$ and $|R_2\rangle$ are mutually orthogonal. We shall compare these states with $|K_1\rangle$ and $K_2\rangle$ and follow a similar approach as neutral kaon physics to establish the desired result. We shall define two non-orthogonal states, similar to $\KS$ and $\KL$, which contains the interferance term of the lowest order diagram and one loop corrections:
  \begin{eqnarray}
  |R_L\rangle =|R_2\rangle+\delta_B|R_1\rangle
  \\
  |R_S\rangle =|R_1\rangle+\delta_B|R_2\rangle .
  \end{eqnarray} 
   Here $R_{L,S}$ is defined as $R_{2,1}$ with a small admixture of $R_{1,2}$, making these states non-orthogonal: $(R_L,R_S)\neq 0$.  Now to construct the BIs, we follow the same approach of Section 3. We consider the states $|R_{L,S}\rangle$, $|R_{1,2}\rangle$ in Hilbert space $\mathcal{H}$. Utilizing the orthogonality conditions for the states $|R_1\rangle$ and $|R_2\rangle$ 
   \begin{eqnarray}
   (R_j,R_k)=\delta_{ij},
   \end{eqnarray}
   where $j=1,2$ and $k=1,2$, and using Eqs. (36) and (37) we express the inner products of the non-orthogonal states as 
   
   \begin{eqnarray}
   \begin{aligned}[b]
   (R_L,R_S)=2Re(\delta_B) \\
   (R_1,R_S)=1=(R_2,R_L) \\
   (R_1,R_L)=\delta_B=(R_2,R_S)
   \end{aligned}
   \end{eqnarray}
  Using these inner products we can now construct the BIs as
 \begin{eqnarray}
 \Delta_3(R_L, R_S, R_1)=2\delta_B Re(\delta_B)\\
 \Delta_3(R_L, R_S, R_2)=2\delta_B^* Re(\delta_B)
 \end{eqnarray}
 Combining these two results we get the following connection 
 \begin{eqnarray}
  \Delta_3(R_L, R_S, R_1)- \Delta_3(R_L, R_S, R_2)=4iRe(\delta_B)Im(\delta_B),
 \end{eqnarray}
making,
\begin{eqnarray}
 \Delta_3(R_L, R_S, R_1)- \Delta_3(R_L, R_S, R_2)\propto \Delta B,
\end{eqnarray}
according to Eq. 35. This result is comparable with Eq. 24 from section [3]. We notice that as we followed the same approach of section [3], the order of the BI appearing in this section is also 3. This result shows the connection between the geometrical BIs with the baryon asymmetry. As according to Sakhrov's condition any successful theory of Baryogenesis must incorporate \CP violation, Eq. 43 also establishes a connection between BIs and \CP violation in the case of Baryogenesis.
Just like section 3, here we constructed third order BIs as fourth order BIs generates no unique results. Using Eqs. 36, 37 and 39 we can construct the following non-zero fourth order Bargmann Invariants 
\begin{eqnarray}
\Delta_4(R_L,R_1,R_S,R_2)=(\delta_B^*)^2 \\
\Delta_4(R_L,R_2,R_S,R_1)=(\delta_B)^2
\end{eqnarray}
With these we obtain a linear relationship between fourth and third order BIs:
\begin{eqnarray}
\begin{aligned}[b]
\Delta_4(R_L,R_2,R_S,R_1)-\Delta(R_L,R_1,R_S,R_2)
&=\Delta_3(R_L,R_S,R_1)-\Delta(R_L,R_S,R_2)
\end{aligned}
\end{eqnarray}
Hence it is sufficient to work only with the third order BIs in this section.

Inspite of the similarity between the ($K_L, K_S$) system with ($R_L, R_S$),
there is one extremely important difference, that there are no self-energy diagram in the 
baryogenesis model we studied. As a result there is no resonant production of 
baryon asymmetry in this model. 

\section{Leptogenesis with right-handed neutrinos}
The observed baryon asymmetry of the universe could be explained from the mechanism 
which generates a lepton asymmetry of the universe before the electroweak phase transition, 
called leptogenesis \cite{lepto}. In the see-saw mechanism for neutrino masses \cite{seesaw}, 
an extension of the standard model is considered where we include heavy right handed Majorana 
neutrinos ($N_{Ri},i=e,\mu,\tau$). The decay of this heavy Majorana neutrinos generates a lepton 
asymmetry, which results into the baryon asymmetry. In this section, we shall present a breif 
reviw of leptogenesis and we shall also present a comparision of this with the case of 
\CP violation in neutral kaon decay.
 The interaction for right handed neutrinos is represented by 
\begin{eqnarray}
\mathcal{L}=h_{\alpha i}\overline{l_L\alpha}\phi N_{Ri}+M_i\overline {N_{Ri}^C}N_{Ri},
\end{eqnarray}
where $l_L\alpha$ are the leptons with $\alpha =1,2,3$ for three generations of leptons, $\phi^T=(-\overline{\phi^0},\phi^-)$ is the Higgs doublet of the standard model and $h_{\alpha i}$ are the Yukawa coplings. The decays of the right-handed neutrinos which violates lepton number:
\begin{eqnarray}
N_{Rj}\rightarrow l_{k}+\overline{\phi}\nonumber \\
\rightarrow l^c_k+\phi
\end{eqnarray}
can generate a lepton asymmetry, provided that there is enough \CP violation.

\CP violation has two independent sources for the case of leptogenesis \cite{Sarkar}. One comes from the interference of vertex and tree-level diagrams in the decays of heavy neutrinos \cite{Fukugita}, which is comparable to the \CP violation coming from the penguin diagram in K decays. Hence we shall call it  $\epsilon^\prime$-type (direct \CP violation). Another source of \CP violation comes from the fact that the heavy neutrino physical states are not \CP or lepton number eigenstates.The physical heavy neutrino states are an admixture of $N_{Ri}$ and $N_{Ri}^C$, and when these physical states are formed, they could decay into both leptons and antileptons, violating \CP \cite{flanz}. This process is comparable to $K-\overline{K}$ mixing and we shall call it $\epsilon$-type (indirect \CP-violation). 

\begin{figure}[h!]
	\centering
	\begin{subfigure}[b]{0.4\textwidth}
		\centering
		\includegraphics[width=\textwidth]{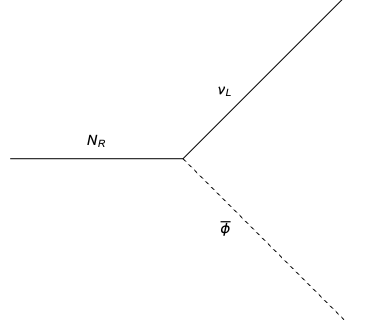}
		\caption{}
		\label{}
	\end{subfigure}
	\hfill
	\begin{subfigure}[b]{0.4\textwidth}
		\centering
		\includegraphics[width=\textwidth]{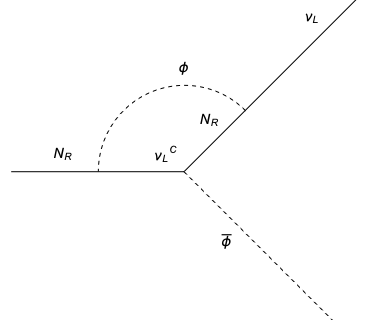}
		\caption{}
		\label{}
	\end{subfigure}
		\caption{Tree level and one loop vertex diagram for the decay of right handed heavy neutrino}
	\label{neutrino_decay}
\end{figure}

The tree level and one loop diagram for the first case, that is the $\epsilon^\prime$-type case, is shown in fig(\ref{neutrino_decay}).

The interference of these two diagrams gives the \CP violating parameter $ \epsilon^\prime$ and we have the following relation
\begin{eqnarray}
\epsilon^\prime \propto Im[\sum_\alpha(h_{\alpha 1}^*h_{\alpha 2}\sum_\beta(h_{\beta 1}^*h_{\beta 2}))]
\end{eqnarray}
This result is for two generations with the limit of large mass difference between $M_1$ and $M_2$.

In the second case, \CP violation enters through the mixing of Majorana mass matrix. This $\epsilon$-type process can be comparable with the mixing of $K-\overline{K}$. The physical state in case of neutral Kaons is $K_S$ and $K_L$, which are the admixtures of the mass eigenstates $K^0$, $\overline{K^0}$ or the \CP eigenstates $K_1$,$K_2$. The linear combination connecting $K_S$, $K_L$ to $K^0$,$\overline{K^0}$(or $K_1$,$K_2$) is assymetric, meaning that the probability of finding $K^0$ ($K_1$) is not the same as the probability of finding $\overline{K^0}$($K_2$). 
When these physical states decay to generate electrons/positrons, this asymmetry survives which gives rise to \CP- violation. 

Majorana particles are their own antiparticles, but when \CP violation is present, we represents the particle and antiparticle independently, for convenience. Representing the right handed Majorana neutrinos as $N_{Ri}$ and the antiparticle as $N_{Ri}^C$, we can write down the interaction Lagrangian as 
\begin{eqnarray}
\mathcal{L}=\sum_i M_i[\overline{(N_{Ri})^C}N_{Ri}+\overline{N_{Ri}}(N_{Ri})^C] \nonumber
\\ 
+\sum_{\alpha,i}h_{\alpha i}^*\overline{N_{Ri}}\phi^\dagger l_{L\alpha}+\sum_{\alpha,i}h_{\alpha i}\overline{l_{L\alpha}}\phi N_{Ri}\nonumber
\\
+\sum_{\alpha,i}h_{\alpha i}^*\overline{l_{L\alpha}^C}\phi (N_{Ri})^C+\sum_{\alpha,i} h_{\alpha,i}\overline{(N_{Ri})^C}\phi^\dagger(l_{L\alpha})^C
\end{eqnarray}

The states $N_{Ri}$ and $N_{Ri}^C$ are analogus to the states $K^0$ and $\overline{K^0}$. $N_{Ri}$ decays only into leptons and $N_{Ri}^C$ decays only into antileptons. Now in the presence of Yukawa interactions one loop corrections to the Majorana mass matrix are found, whose eigenfunctions are an admixture of $N_{Ri}$ and $N_{Ri}^C$. These physical states were shown to be asymmetrical linear combinations of $N_{Ri}$ and $N_{Ri}^C$, giving rise to a \CP  asymmetric universe. 

\begin{figure}[h!]
	\centering
	\begin{subfigure}[b]{0.4\textwidth}
		\centering
		\includegraphics[width=\textwidth]{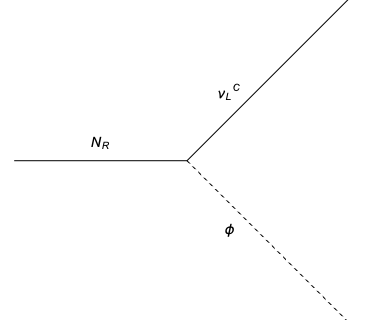}
		\caption{}
		\label{}
	\end{subfigure}
	\hfill
	\begin{subfigure}[b]{0.4\textwidth}
		\centering
		 \includegraphics[width=\textwidth]{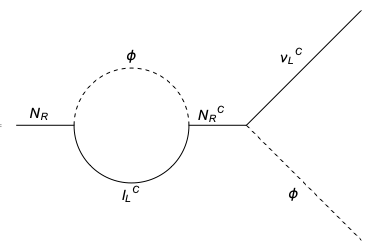}
		\caption{}
		\label{}
	\end{subfigure}
	
	\caption{(a)Tree level and (b)one loop self-energy diagrams for right handed neutrinos }
	\label{neutrino_decay_corr}
\end{figure}

Fig (\ref{neutrino_decay_corr}) shows the tree level and one loop 
self-energy corrections for right handed neutrinos giving rise to lepton asymmetry. 

The \CP violation, which enters through the non-diagonal mass matrix is also seen to be proportional to 
\begin{eqnarray}
\delta \propto Im[\sum_\alpha(h_{\alpha 1}^*h_{\alpha 2}\sum_\beta(h_{\beta 1}^*h_{\beta 2}))]
\end{eqnarray},
with a different phase factor than $\epsilon^\prime$.

\section{Bargmann invariants in the theory of Triplet Higgs Leptogenesis }
\begin{figure}[h!]
	\centering
	\begin{subfigure}[b]{0.4\textwidth}
		\centering
		\includegraphics[width=\textwidth]{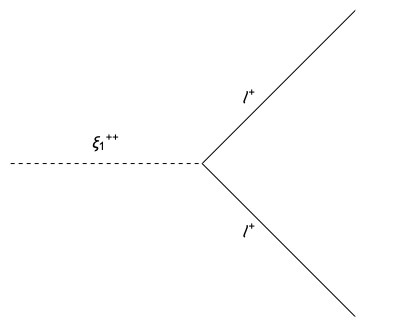}
		\caption{}
		\label{}
	\end{subfigure}
	\hfill
	\begin{subfigure}[b]{0.45\textwidth}
		\centering
		\includegraphics[width=\textwidth]{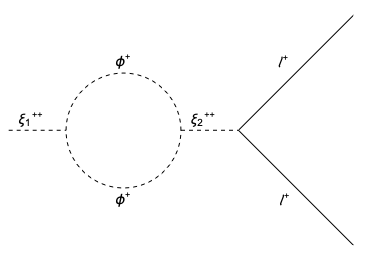}
		\caption{}
		\label{}
	\end{subfigure}
		\caption{The decay of $\xi_1^{++}$ at(a) tree level and in (b) one loop diagram }
	\label{triplet_higgs}
\end{figure}

In this section we shall discuss leptogenesis in the triplet Higgs mechanism of neutrino masses \cite{MaSarkar} and develop a connection between the lepton asymmetry and the BIs. The interactions of the heavy Higgs triplets ($\xi_a$, a=1,2) relevant for the study of generating leption number asymmetry is given by:
\begin{eqnarray}
\mathcal{L}=f_{aij}\xi_al_il_j+\mu_a\xi_a^{\dagger}\phi\phi,
\end{eqnarray}
where $f_{aij}$ are the Yukawa couplings, $\mu_a$ are the chemical potentials. The lepton-number violating decays that can account for the lepton asymmetry of the universe, if the interactions are slow enough and there is enough \CP violation, is given by:
\begin{eqnarray}
\xi_a^{++}\rightarrow l_j^+l_k^+  \qquad (L=2),\nonumber \\
\xi_a^{++}\rightarrow \phi^+\phi^+ \qquad (L=0).
\end{eqnarray}
The simultaneous existance of these two decay processes indicates an asymmetry in the lepton numbers. The mass matrix of $\xi_a$ at the tree level can be expressed as diagonal and real, without the loss of any generality, meaning \CP is conservation at this level. But \CP non-conservation may occur due to the interactions between tree and one-loop diagrams, as shown in fig (\ref{triplet_higgs}). Note that here we do not have any vertex corrections that can incorporate \CP violation, unlike other models of baryogenesis and leptogenesis discussed before. 

In the one-loop correction level the mass matrices become non-diagonal, introducing \CP non-conjugate physical states. A detailed calculation, with the consideration that $\xi_2$
is heavier than $\xi_1$, generates a lepton asymmetry:
 \begin{eqnarray}
 \delta_L={Im[\mu_1 \mu_2^* \sum_{i,j}f_{1ij}f_{2ij}^*]\over 8\pi^2(M_1^2-M_2^2)} {M_1\over \Gamma_1}
 \end{eqnarray}
 
 We shall now calculate the BIs for this system and establish a connection between the BIs and lepton asymmetry for the triplet Higgs leptogenesis. To do so, let us consider a Hilbert space $\mathcal{H}$ which contains the states $\xi_a$, $l_j$ and $\phi$. We can construct the BI from the Lagrangian of triplet Higgs theory following the same approach we used for neutral Kaon decay. 
 To construct the necessary inner products, we consider the following relations
 \begin{eqnarray}
 \xi_a=f_{aij}l_il_j \nonumber\\
 \xi_a^\dagger =\mu_a\phi\phi.
 \end{eqnarray}
 Now we can express the inner products as
 \begin{eqnarray}
 (\xi_a,l_il_j)=f^*_{1ij}\nonumber \\
 (\xi_a,\phi\phi)=\mu_a
 \end{eqnarray}
 We now construct the following BI:
 \begin{eqnarray}
 \begin{aligned}[b]
 \Delta_4(\xi_1,\phi\phi,\xi_2,ll)
 &=(\xi_1,\phi\phi)(\phi\phi,\xi_2)(\xi_2,ll)(ll,\xi_1) \\
 &=Tr[\rho(\xi_1)\rho(\phi\phi)\rho(\xi_2)\rho(ll)] \\
 &=\mu_1\mu_2^*f_{2ij}^*f_{1ij}
 \end{aligned}
 \end{eqnarray}
 So, we can visualize the above BI as paths in the ray space connecting the points $\rho(\xi_1)$ to $\rho(\phi\phi)$ and so on, ultimately coming back to point $\rho(\xi_1)$ making a closed loop. These ray space points can be any lift of $\mathcal{B}$, making the BI rephasing invariant. This BI physically signifies the interaction between the tree level and the one loop diagram of $\xi_1^{++}$ decay. Now, the BI we just obtained appears in the expression of $\delta_L$ and are connected as:
 \begin{eqnarray}
 \delta_L=Im( \Delta_4(\xi_1,\phi\phi,\xi_2,ll))
 \end{eqnarray}
 
 This connects the geometrical BI with the physical lepton asymmetry for the triplet Higgs theory of leptogenesis. The BI we obtained gives the geometrical phase for the theory:
 \begin{eqnarray}
 \phi_g=-arg(\Delta_4).
 \end{eqnarray}
 The similarity of approaching the problem between this section and section [4] leads to the appearence of fourth order BI, just as it appeared in section [4].

\section{Summary}
In this paper we have presented the connection between the Bargmann invariants and rephasing invariant quantities for neutral Kaon oscillation, Kaon decays, Baryogenesis, leptogenesis with right-handed neutrinos and leptogenesis with triplet Higgs scalars. These results also connect the rephasing invariant quantities to the geometric phases. In each case we have expressed the connection between \CP violation and BIs. 

We studied the $K^\circ - \bar{K^\circ }$ oscillation and kaon decaying 
to pions with mesons as the effective fields, without resorting to the
quark model. The interference of tree level and one-loop self-energy and vertex corrections were considered in the effective meson theory to conect CP violation and BIs, and hence the geometric phases. Self energy corrections were analyzed for the case of neutral Kaon oscillation ($\Delta S=2$) and triplet Higgs Leptogenesis and vertex corrections were analyzed for the case of Kaon decay ($\Delta S=1$) and Baryogenesis to establish such connections.
 
For the case of neutral Kaon system, we can get \CP violation from both self energy correction and vertex corrections. Hence the Bargmann Invariants were shown to be connecting rephasing invariant parameter for both the cases. But for the $XY$ model of Baryogenesis \CP violation arrises only from vertex correction and for the triplet Higgs theory of Leptogenesis \CP violation arises only from the self energy correction. So, BIs were shown to be connected with rephasing invariants for those respective cases only. In the process of connecting rephasing invariants with the geometric phases third order Bargmann Invariant appears in self energy correction of neutral Kaon system and in vertex diagram of Baryogenesis; while fourth order Bargmann Invariant appears in vertex correction for Kaon decay and in self energy diagram of triplet Higgs Leptogenesis. The existance of a linear relationship between third and fourth order BIs for the first two cases stated above explains why it is sufficient to only work with the third order BIs.

{\it Acknowledgement :} We would like to thank Prof. Arghya 
Taraphder for encouragement. 
The work of U.S.  was supported by 
a research grant associated with the Raja Ramanna  Fellowship of DAE, India.

\newpage

\baselineskip 18pt
\bibliographystyle{unsrt}

\end{document}